\documentclass[aps,prx,reprint,twocolumn,floatfix,superscriptaddress,amsmath,amssymb,amsfonts,nofootinbib,balancelastpage]{revtex4-2}
\usepackage[colorlinks=true, linkcolor=blue, citecolor=blue, urlcolor=blue, allcolors=blue]{hyperref}

\usepackage[left=1.5cm,right=1.5cm,top=2cm,bottom=2cm]{geometry}
\usepackage{xcolor}
\usepackage{graphicx,amsmath,wasysym}% Include figure files
\usepackage{dcolumn}% Align table columns on decimal point
\usepackage{bm}% bold math

\usepackage{natbib} % Load natbib here
\usepackage{geometry}
\usepackage{ifthen}
\usepackage{slashed}
\usepackage{graphicx}

\usepackage[colorlinks=true,linkcolor=blue,citecolor=blue,urlcolor=blue,allcolors=blue]{hyperref}  % Enable hyperlinking

\usepackage{amssymb}

\makeatletter
\def\@tvsp{\mathchoice{{}\mkern-4.5mu}{{}\mkern-4.5mu}{{}\mkern-2.5mu}{}}
\def\ltrivert{\left|\@tvsp\left|\@tvsp\left|}
\def\rtrivert{\right|\@tvsp\right|\@tvsp\right|}
\makeatother
\usepackage{amssymb}
\begin{document}
\title{Microscopic theory of field-tuned topological transitions in the Kitaev honeycomb model}% Force line breaks with \\

\author{Jagannath Das}

\email{jagannath.das@theory.tifr.res.in}
\affiliation{Department of Theoretical Physics, Tata Institute of
Fundamental Research, Homi Bhabha Road, Colaba, Mumbai 400005, India.}
\author{Aman Kumar}
\email{akumar@magnet.fsu.edu}
\affiliation{ National High Magnetic Field Laboratory, Tallahassee, Florida 32310, USA}
\affiliation{Department of Physics, Florida State University, Tallahassee, Florida 32306, USA}
\author{Vikram Tripathi}

\email{vtripathi@theory.tifr.res.in}
\affiliation{Department of Theoretical Physics, Tata Institute of
Fundamental Research, Homi Bhabha Road, Colaba, Mumbai 400005, India.}


\begin{abstract}
\iffalse
Topological phases and transitions in the Zeeman-perturbed isotropic Kitaev model have remained open questions primarily because the crucial effect of gauge fluctuations is not properly included in mean field treatments. \fi
We microscopically construct an abelian mutual Chern-Simons lattice gauge theory for magnetic field-tuned topological transitions in the Kitaev model and obtain a complete characterization of the phases, including their quasiparticles. At low fields for both ferro and antiferromagnetic (FM(AFM)) Kitaev interactions, we demonstrate nonabelian Ising topological order (ITO), and explicitly construct the Majorana anyon as an intrinsic excitation -- a twist defect in our \textit{abelian} gauge theory. For the AFM case, an abelian chiral phase appears at intermediate fields with trivial topological order and fermionic bulk excitations. Remarkably, both the ITO phase and the intermediate phase have the same chiral central charge $c=1/2,$ implying no change in the quantized thermal Hall response across the transition. For the FM case, there is a direct transition from ITO to a partially polarized nontopological phase. Our study completes the proof of Kitaev's original proposal of the low-field ITO with $c=1/2$ going beyond his mean-field arguments by including the crucial effect of the gauge fluctuations, and provides a resolution of the debate surrounding the intermediate field phase in the AFM case.
\end{abstract}
\maketitle

Study of topological orders and their stability against perturbations is a central topic of current interest in condensed matter physics \cite{anderson1987resonating,baskaran1993resonating,kalmeyer1987equivalence,kivelson1987topology,affleck1988large, read1991large, wen1989vacuum,wegner1971duality,senthil2000z,kogut1979introduction,moessner2001resonating,kitaev2006anyons,savary2016quantum}. 
It was originally proposed by Kitaev \cite{kitaev2006anyons}, using a mapping to a mean-field chiral $p$-wave superconductor ($p$SC) model, that the Zeeman-perturbed isotropic Kitaev model has nonabelian Ising topological order (ITO) because (a) Majorana zero modes associated with \textit{extrinsic} twist defects ($\pi$-flux particles) of the $p$SC obeyed Ising braiding rules, and (b) the chiral central charge of the mean field model was $c=1/2.$ However the effect of the gauge sector must be included to demonstrate the nonabelian Majorana anyon as an \textit{intrinsic} bulk excitation, and also obtain the correct values of $c,$ topological entanglement entropy $\gamma,$ and ground state degeneracy (GSD). For mean field  $p$SC models, it is long known that three or four different ground states on the torus are possible \cite{read_green2000}; however, such ground states are unique and do not simultaneously belong to a degenerate ground state sector. Specifically, the three-fold degenerate ITO state in this case is known to require nonabelian gauge fluctuations \cite{read_green2000} while in Kitaev's fermionization scheme, the gauge fields that appear naturally are abelian $Z_2$. In the low field regime, recent perturbative treatments \cite{joy2022dynamics,chen2023nature} emphasize the important role of vison (gauge field) fluctuation effects, and numerical studies suggest a low-field ITO phase \cite{gohlke2018dynamical, joy2022dynamics, chen2023nature, kumar2023thermal,hickey2019emergence} but the understanding of quasiparticles, necessary for complete characterization of the topological phases, is lacking. In the AFM case there is an ongoing debate regarding the nature of the  intermediate field phase \cite{zhu2018robust,teng2020unquantized,feng2023dimensional,nasu2018successive,liang2018intermediate,gohlke2018dynamical,patel2019magnetic,hickey2019emergence,hickey2021generic,zhang2022theory,jiang2020tuning,das2024field}. The high field phase is topologically trivial and partially polarized. For the FM  Kitaev model, there is a direct field-tuned transition from the  Kitaev spin liquid (KSL) to the partially polarized phase.

\iffalse
At high fields, the partially polarized phase has interesting features such as dimensional reduction of the excitations, which has been understood using spin waves \cite{joshi2018topological} as well as gauge theory approaches \cite{das2023jordan}. 
\fi
Experimentally, there is great interest in the possibility of ITO revival in Kitaev materials near field-suppressed magnetic order \cite{yokoi2021half,bruin2022robustness}. Such questions have also inspired phenomenological studies of topological phases and excitations in relevant matter Chern-Simons field theories \cite{nayak2008non,zou2020field}. However before attempting a microscopic treatment of field-tuning ITO in Kitaev materials, it is important to first understand the behavior in the Kitaev limit.  

We present here a microscopic derivation and analysis of an effective topological gauge theory for the isotropic Kitaev model subjected to a $(001)$ Zeeman and (Haldane mass) three-spin perturbations. We use a recent Chern-Simons spin fermionization scheme \cite{das2023jordan} valid for arbitrary $2D$ lattices that avoids enlargement of the local Hilbert space, with the fermions attached in specific ways to two abelian $Z_2$ gauge fields respectively on the lattice and dual lattice.
\iffalse Such enlargement of the gauge degrees of freedom is necessary for satisfaction of the Chern-Simons Gauss-law constraints on arbitrary lattices generally lacking local one-to-one face-vertex correspondence \cite{sen2000pre,susskind1991,banks2021,kapustin}.  We construct the quasiparticles and obtain their fusion rules, together with topological ground state degeneracy (GSD), $\gamma,$ and $c$ in different field regimes of the FM and AFM Kitaev interactions. \fi 
Our analysis gives the first complete characterization of low-field ITO as well as intermediate field phases in the AFM Kitaev model  starting from a microscopic theory.

Our main results are as follows (see  Table \ref{fig:table}). At low fields, in both FM and AFM cases, the fusion rules, 3-fold GSD on the torus,  $\gamma=\ln 2$ and $c=1/2$ substantiate the phase has ITO, proving Kitaev's original proposal. For the FM case, ITO is lost beyond a critical field $h_c$ and the transition is to a topologically trivial partially polarized state characterized by $\gamma=c=0.$ For the AFM case, the system transitions at $h_{c1}$ (appreciably larger than $h_c$ of the corresponding FM case) to a chiral abelian phase with fermion quasiparticles. This phase which persists at intermediate fields $h_{c1}<h<h_{c2}$ has trivial topological order i.e. $\gamma=0,$ no GSD, and $c=1/2,$ and belongs to the class of a $p+ip$ superconductor where fermion number parity is not gauged. Significantly, \emph{none} of these facts about the intermediate phase can be captured by parton mean field theory, and neglecting gauge fluctuations leads to erroneous results such as sign reversal of half-quantized thermal Hall response \cite{nasu2018successive} at $h_{c1}.$ Beyond  $h_{c2},$ the partially polarized phase appears like the FM case.  
\begin{table}
\centering
\includegraphics[width=\columnwidth]{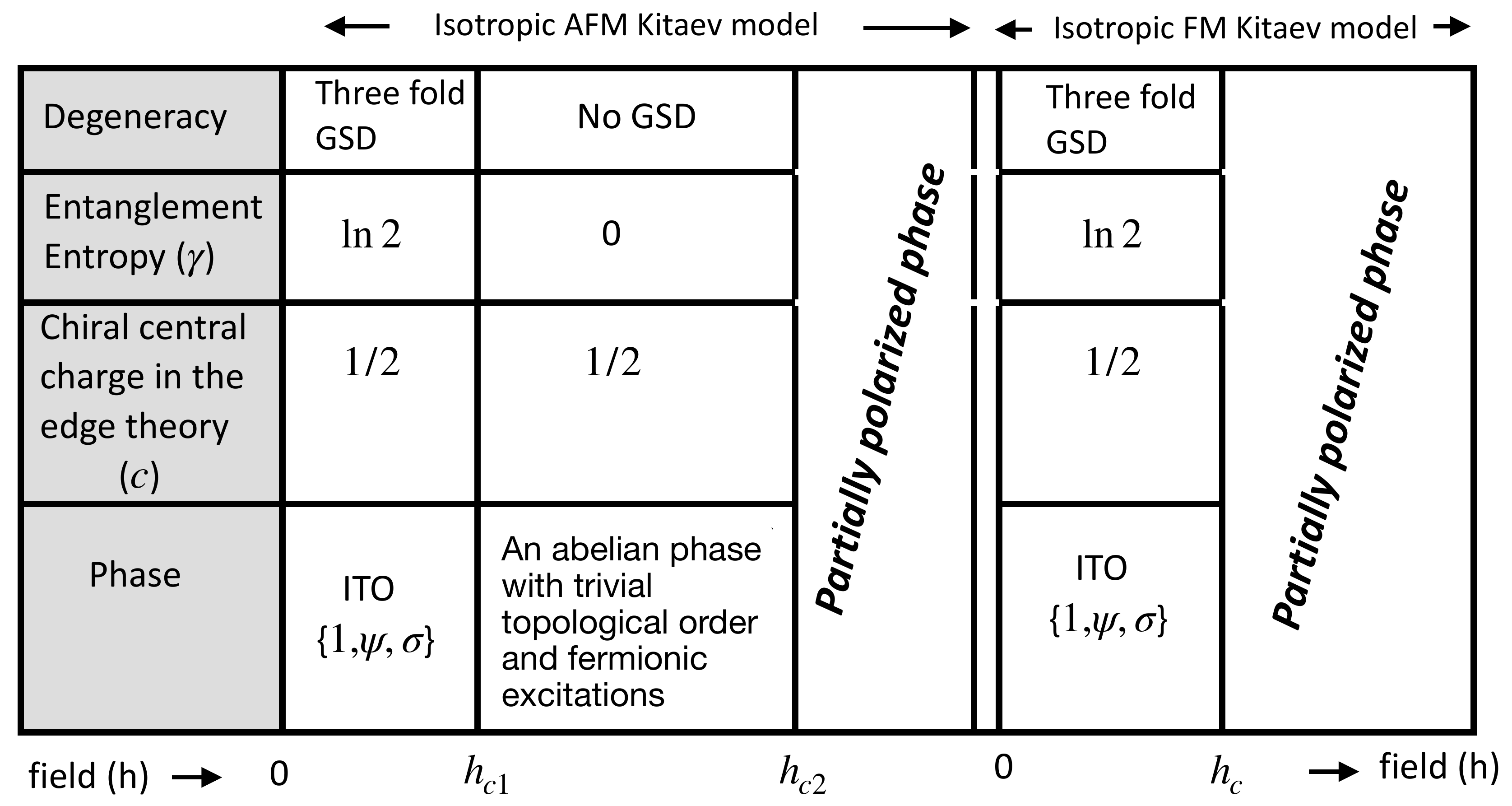}
\caption{\label{fig:table} The table gives a complete characterization of the different field-tuned topological phases of  both  the FM and AFM Kitaev model.}
\end{table}

We begin with the following $SO(3)$ Majorana fermionization of the spin-1/2 ($\mathbf{S}=\pmb{\sigma}/2$) \cite{tsvelik1992new},
\begin{align}
\sigma^{x}_{i}=-i\eta^{y}_{i}\eta^{z}_{i},\hspace{0.25in}
\sigma^{y}_{i}=-i\eta^{z}_{i}\eta^{x}_{i}, \hspace{0.25in}
\sigma^{z}_{i}=-i\eta^{x}_{i}\eta^{y}_{i}.
\label{SO(3)}
\end{align}
The singlet operators $\gamma_i=-i\eta^x_i\eta^y_i\eta^z_i,$ anticommute with all fermions except with the $\eta_i$ fermions at the same site $i,$ and act as Jordan-Wigner (JW) strings \cite{fu2018properties,fu2018three}. \iffalse and the 2D Jordan-Wigner (JW) transformation provides a nonlocal and nonlinear mapping between the spin and fermion descriptions \cite{Kamenevprb2020,kamenev2017prb,lopez1994chern,kumar2014chern}. \fi Field theoretically, the 2D JW transformation \cite{Kamenevprb2020,kamenev2017prb,lopez1994chern,kumar2014chern} is implemented by Chern-Simons (CS) flux attachment \cite{fradkin1989prl,wang1991ground,azzouz1993interchain,derzhko2001jordan}, but on the lattice, consistency is achieved only if local one to one face-vertex correspondence is satisfied \cite{fradkin2015prb}. However enlarging the gauge fields by including the dual lattice allows construction of consistent mutual CS theories on arbitrary 2D lattices  \cite{banks2021,susskind1991,sen2000pre,kapustin} with Lagrangian (see  Supplemental Material (SM) \cite{supp.} for details)
\begin{align}
\mathcal{L}_{\text{CS}} & =\!\!\frac{k}{4\pi}[\xi^{*}_{f^{*}e^{*}} a^{*}_{e^{*}}A_{v}\delta_{f^{*}v}+\!D^{*}_{v^{*}e^{*}}a^{*}_{v^{*}}A_{e} \delta_{ee^{*}}-\partial_{0}a^{*}_{e^{*}}A_{e}\delta_{ee^{*}}] \nonumber \\
&\!\!\!\!\!\!\!\! -\frac{k}{4\pi}[\xi_{fe} A_{e} a^{*}_{v^{*}}\delta_{fv^{*}}+D_{ve}A_{v}a^{*}_{e^{*}} \delta_{ee^{*}}-\partial_{0}A_{e}a^{*}_{e^{*}}\delta_{ee^{*}}],
\label{eq:mixed_CS}
\end{align}
where $\xi_{fe}$ and $D_{ve}$ are respectively the lattice analogs of curl and gradient \cite{fradkin2015prb} operations and $``*"$ everywhere labels the dual lattice. The $\delta$-functions are defined as follows: $\delta_{e,e^{*}} = 1$ if $e$ and $e^{*}$ are links dual to each other, and zero otherwise, and similarly for $\delta_{fv^{*}}$ etc. 
Hereafter we make the gauge choice that all the temporal components of the electromagnetic fields are zero, i.e. $A_v = a^{*}_{v^{*}}=0.$  We fermionize the spin operators,
$
\sigma^{\alpha}_{i}=\gamma_i \eta_{i}^{\alpha},$
using the JW string, $\gamma_i=e^{i\int^i_{\mathcal{C}} (B_{e}+A_{e})},$ where
$B_e=\frac{k}{2}(a^{*}_{e1^{*}}+a^{*}_{e2^{*}}+a^{*}_{e3^{*}}+a^{*}_{e4^{*}})$ is a specific combination of dual lattice  gauge fields \cite{das2023jordan} enclosing the link $e,$ chosen to satisfy the spin commutation relations \footnote{The normalization in $B_e$ is $\frac{k}{2}$ which differs from Ref. \cite{das2023jordan} by a factor of two, and corrects it.}. $\mathcal{C}$ denotes a path that starts from some reference point and ends at lattice site $i.$  We choose both $A_{e}$ and $a^{*}_{e^{*}}$ as $Z_{2},$ so gauge invariance constraints models we study to those conserving fermion and vortex number parity. From the property $\sigma^2_i=1,$ we additionally require $ e^{i\int_e 2B^{i}_{e}}=1$. For $Z_2$ gauge fields, 2-body spin interactions on any link $e= (ij)$ have a simple form,
\begin{align}
    \sigma^{\alpha}_i\sigma^{\beta}_j=\eta^{\alpha}_i e^{i{(A_e+B_e)}} \eta^{\beta}_j.
    \label{spin_com}
\end{align}
Comparing with the hermitian conjugate of  Eq. \ref{spin_com} and using the anticommutation property of the $\eta$-fermions, we obtain the constraint $e^{iB_e}=e^{i\frac{(2n+1)\pi}{2}}$ with $n$ an integer. The conditions $e^{2iB_{e}}=-1$ and $e^{i\int_{e} 2B_{e}}=1$ ensure the correct spin statistics \cite{supp.}.
As a corollary of Eq. \ref{spin_com}, 
for every link $e=\langle ij\rangle,$ $\langle \gamma_i \gamma_j \rangle = e^{i(A_e+B_e)}.$
We choose the level $k=1$ which describes a trivial state \cite{das2023jordan}.

We now proceed to implement Chern-Simons fermionization of the isotropic Kitaev model,
\begin{align}
    \mathcal{H}_{K}=-K\sum_{\langle pq\rangle \in \vartheta-\rm{links}}\sigma_{p}^{\vartheta}\sigma_{q}^{\vartheta}.
    \label{eq:Kitaev}
\end{align}
Here $p,q$ are the  vertices associated with the corresponding link $\vartheta,$ with $\vartheta = x,\,y,\,\mbox{or }z.$ 
Using the $SO(3)$ representation of Eq. \ref{SO(3)} we write
$
\sigma^{x}_{i} \sigma^{x}_{j}  = i(-i\eta^y_i\eta^y_j)\eta^z_i\eta^z_j =ie^{iA^x_e}\eta^z_i\eta^z_j$ and  
$ \sigma^{y}_{i} \sigma^{y}_{j} =i(-i\eta^x_i\eta^x_j)\eta^z_i\eta^z_j =ie^{iA^y_e}\eta^z_i\eta^z_j, $
where $e^{iA^x_e}=-i \eta^{y}_{i}\eta^{y}_{j}$ for the $x-$link and $e^{iA^y_e}=-i \eta^{x}_{i}\eta^{x}_{j}$ for the $y-$link are chosen such that commutation relations of these $\eta$ bilinears are preserved. We perform a Hubbard-Stratonovich decoupling of the four-fermion interaction on the $z-$bond following Ref. \cite{nasu2018successive}, 
$-\sigma^{z}_{i} \sigma^{z}_{j} =  -(\gamma_i\eta^z_i)(\gamma_j\eta^z_j)$ $ = - m_i \gamma_j\eta^z_j - m_j \gamma_i\eta^z_i $ $+ m_i m_j -i\delta \eta^z_i\eta^z_j$ $- \gamma_i i\gamma_j \Delta + \delta \Delta,$
with the mean fields $m_i=\langle \gamma_i \eta^z_i \rangle,$  $\delta=\langle i\gamma_i \gamma_j \rangle$ and $\Delta=\langle i\eta^z_i \eta^z_j \rangle.$  We now introduce  the following time-reversal symmetry breaking terms that preserve fermion number parity:  $\mathcal{H}_Z=-h\sum_i \sigma^z_i,$ is a Zeeman $(001)$ perturbation and $ \mathcal{H}_{\kappa}=-\kappa\sum_{j,k,l} \sigma^{x}_{j}\sigma^{y}_{k} \sigma^{z}_{l},$ corresponds to the scalar spin chirality (Haldane mass term).
\iffalse
 \begin{align}
     \mathcal{H}_Z + \mathcal{H}_{\kappa} & =-h\sum_i \sigma^z_i -\kappa\sum_{j,k,l} \sigma^{x}_{j}\sigma^{y}_{k} \sigma^{z}_{l},
 \end{align}
 where $\mathcal{H}_Z$ is a Zeeman perturbation, and $\mathcal{H}_{\kappa}$ corresponds to the scalar spin chirality (Haldane mass term).
\fi
The $(001)$ orientation has garnered interest recently because of enhanced field-robustness of the KSL \cite{nasu2018successive,feng2023dimensional}. For small deviations of the field direction from $(001),$ the Haldane mass term -- necessary for chiral phases  -- is also generated. The model now has two parameters $h$ and $\kappa;$ however, $\mathcal{H}_{\kappa}$ commutes with the vison (flux) operators and therefore will not cause the vison fluctuations necessary to bring about the topological transitions; also see SM \cite{supp.}.
The mean field Hamiltonian in momentum space is \cite{nasu2018successive} (see SM \cite{supp.} for details)
\begin{align}
\mathcal{H}_K & = \sum_{\vec{k}}\Psi^{\dagger}_{\vec{k}} \mathcal{M}_K \Psi_{\vec{k}} +NK_z \delta \Delta +NK_z m^2,
\label{eq:Hmf}
\end{align}
with $ \Psi^{T}_{\vec{k}}=\begin{bmatrix}
 i\gamma_{\vec{k}A} &
 i\gamma_{\vec{k}B} &
 \eta^{z}_{\vec{k}A} &
 \eta^{z}_{\vec{k}B} 
 \end{bmatrix}$
and
\begin{equation}
 \mathcal{M}_{K}=
 \begin{bmatrix}
 0 & \frac{i{K}{\Delta}}{2} & -\frac{i({Km+h})}{2} & 0 \\ 
- \frac{i{K}{\Delta}}{2} & 0  & 0 & -\frac{i({Km+h})}{2} \\ 
\frac{i({Km+h})}{2} & 0 & \frac{W}{2}  & \frac{-iKf}{2} \\ 
 0 & \frac{i({Km+h})}{2} & \frac{iKf^{*}}{2}& -\frac{W}{2} \\ 
 \end{bmatrix},
 \label{Eq:hamilt1}
 \end{equation}
  where  $W^2=\kappa^2 \{
  \delta \sin(0.5k_x + 0.5\sqrt{3}k_y)+ 
  \delta \sin(0.5k_x - 0.5\sqrt{3}k_y) - \sin(k_x)\}^2$ and $f=e^{-ik_{1}}+e^{-ik_{2}}+\delta,$ and $m_{i_A}=m_{i_B}=m.$
$N$ is the number of unit cells, $A$ and $B$ are sublattice indices, and $k_{1}=\Vec{k}.\hat{n}_{1}$, $k_{2}=\Vec{k}.\hat{n}_{2}$ where $\hat{n}_{1}$ and  $\hat{n}_{2}$  are  the unit vector along the $x$ and $y$-links respectively.
 The mean fields are calculated self-consistently using Eqs. \ref{eq:Hmf} and \ref{Eq:hamilt1}   \cite{nasu2018successive}.
 \iffalse
 from the the effective action,
 \begin{align}
  S_{F}= -\frac{1}{\beta} \sum_{n,\Vec{k},\lambda}\ln (-i\omega_{n}+\lambda^{\Vec{k}})+NK_z( \delta \Delta +m^A_i m^B_i)
  \label{eq:effectiveaction}
 \end{align}
 where $\omega_n$ is the fermionic Matsubara frequency and $\lambda$ runs over the eigenvalues of the matrix in Eq. (\ref{Eq:hamilt1}). 
 \fi
 \iffalse
 The unperturbed Kitaev model has two gapless Dirac-dispersing bands on either side of zero energy (associated with $\eta^z$) while the other two associated with $\gamma$ fermions are flat and gapped and represent the flux excitations. The Zeeman term hybridizes the $\gamma$ and $\eta^z$ modes making the former disperse but remaining non-chiral.  
 \fi
 The low-lying bands are chiral, with the Dirac points located at $\vec{k}_0=(k_0,0),$ where 
\begin{align} 
\cos\frac{k_0}{2} & =-\frac{1}{2\Delta}\{\Delta\delta+(Km+h)^2\}=-\frac{\xi}{2}.
\label{eq:xi}
\end{align}
 \iffalse
\begin{align}
    \cos\frac{k_0}{2}=-\frac{1}{2\Delta}\{\Delta\delta+(Km+h)^2\}=-\frac{\xi}{2}.
    \label{Eq:gap}
\end{align}
\fi
\iffalse
 Expanding the dispersion around $\vec{k}_0$ gives
$
    E_{\vec{k}} \sim  \pm \sqrt{|v_x|^2 \Tilde{k}^2_x+|v_y|^2 \Tilde{k}^2_y}$,
with $\vec{\Tilde{k}}_0=\vec{k}-\vec{k}_0,$  and
$v_x = \text{sgn}(k_0)\frac{K\Delta \sqrt{4-\xi^2}}{4(\delta-\Delta-\xi)},   \hspace{0.2in} v_y=\frac{\sqrt{3}K\Delta \xi}{4(\delta-\Delta-\xi)}.$ \fi
An advantage of this mean field scheme over earlier parton mean field studies, e.g. \cite{zhang2022theory,jiang2020tuning}, is the avoidance of Hilbert space enlargement that can introduce extra topological phases. Most of the numerical studies \cite{patel2019magnetic,gohlke2018dynamical,zhu2018robust} broadly show only two field-tuned transitions like we find here. See  SM \cite{supp.} for a validation of our mean field choice by benchmarking with extreme limit results.

To obtain the effective gauge action, we first integrate out the massive fermions yielding
\begin{align}
    S_{\text{e}}\!=\!-\frac{1}{2}\text{Tr}\ln\{G^{-1}_{0}\!+\!\hat{T}\}\!+\!S_{CS}\!+\!NK_z( \delta \Delta +\! m^2).
    \label{Eq:S}
\end{align}
Here $G^{-1}_{0} = (-i\omega_n + \mathcal{M}_K)$ is the inverse Green function corresponding to the mean-field Hamiltonian (i.e. without the gauge fluctuations). \iffalse We ignore spatial and temporal fluctuations of the amplitudes of the mean fields since the transition to the partially polarized phase at higher fields is known \cite{das2023jordan} to be driven by phase or gauge field fluctuations. \fi The matrix $\hat{T}$ contains the gauge fluctuation parts of both Kitaev and the scalar spin chirality terms \cite{supp.}. 

Expanding the logarithm term in Eq. (\ref{Eq:S}) we get first  non-trivial contributions  coming from the second order for the links terms. For the $x(y)$-bonds  we get Maxwell-type electric field terms,
\iffalse
 Expanding the logarithm term in Eq. (\ref{Eq:S}) we get first  non-trivial link terms in the second order. For example $\text{tr}(G_{0} \hat{T}G_{0} \hat{T})$ for an $x$-bond $i_A\rightarrow j_B$ ;
 \begin{align}
 \text{tr}(G_{0}TG_{0}T)  & =\!\frac{1}{\beta^2 V^2}\!\!\sum_{A,B}^{k_1,k_2,m,n}\!\!\! \int \! \!d\tau d\tau'  e^{-i\omega_{n}(\tau-\tau')}e^{-i\omega_{m}(\tau'-\tau)}
 \nonumber \\
& \times G^{AA}_0(k_n){{\hat{T}}}^{xAB}(\tau')G^{BB}_0(k_m) {{\hat{T}}}^{xBA}(\tau).
 \label{eq:trace1}
 \end{align}
 Making a change of variables $\tau=\tau_c+\frac{\tau_{r}}{2}$ and $\tau'=\tau_c-\frac{\tau_{r}}{2}$ and performing a gradient expansion, 
 \fi
$
\mathcal{L}_{x(y)} = {E_c}^{-1}(\dot{A}^{x(y)}_e)^{2},
$
with $E_c \approx \frac{W}{2} \frac{\left[\Delta^2+(m+\frac{h}{K})^2\right]^2}{(m+\frac{h}{K})^2},$ and for the $z$-bonds, 
$\mathcal{L}_z =\delta^2{E_c}^{-1} (\dot{A}^z_e+\dot{B}^z_e)^2$. As a high field benchmarking we check that $E_c\sim h^3/K^2$ grows rapidly with $h,$ while the order parameter $\delta$ vanishes resulting in an effectively one-dimensional
character of the low-excitations in the $x$-$y$ backbone reported earlier \cite{das2023jordan,feng2023dimensional}.  
Chern-Simons terms arise from processes shown in Fig. \ref{fig:loop_cs}. For example the configuration in Fig. \ref{fig:loop_cs}(a) gives \cite{supp.}:
\begin{figure}
\centering
\includegraphics[width=\columnwidth]{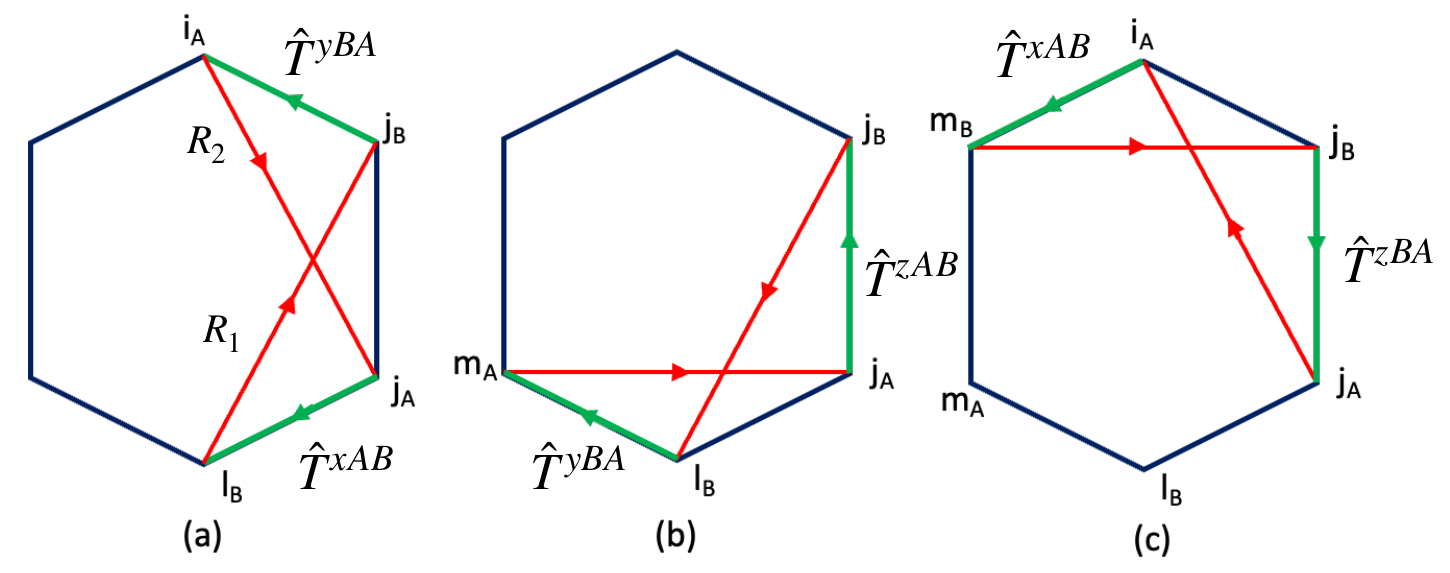}
\caption{\label{fig:loop_cs}The configurations that give  CS terms. The green lines represent the $\hat{T}$ matrices and red lines are associated with Green function elements.}
\end{figure}
\begin{align}
    \Delta S_{CS}=\frac{i}{4\pi} \text{sign}(\xi)\,\text{sign}(W')\sum_{\langle\langle ee'\rangle\rangle}\int d\tau A^x_e \frac{\partial A^y_{e'}}{\partial \tau}
    \label{Eq:usual},
\end{align}
with $e=j_A\rightarrow l_B $ and $e'=i_A\rightarrow j_B. $  It is easy to obtain similar CS contributions from the diagrams in Fig. \ref{fig:loop_cs}b and \ref{fig:loop_cs}c. Note that the two link fields in our CS action are separated by precisely one link and do not share a vertex. 
In our formalism, the $z$-links are associated with the gauge fields $A_e + B_e.$ Consequently, diagrams containing one $z$-link (e.g. Fig. \ref{fig:loop_cs}b, \ref{fig:loop_cs}c) also generate a correction to the mutual CS interaction:
\begin{align}
   \Delta S_{MCS}=\frac{i}{4\pi} \text{sign}(\xi)\,\text{sign}(W')\sum_{\langle\langle ee'\rangle\rangle}\int d\tau A^y_e \frac{\partial B^z_{e'}}{\partial \tau}.
\label{Eq:mutual}
\end{align}
\iffalse The fact that we have explicitly constructed a lattice CS gauge theory for a chiral phase on the honeycomb lattice contradicts the claim in Ref. \cite{fradkin2015prb} that this is possible only for lattices with one-to-one face vertex correspondence. \fi \iffalse However our construct is consistent with a no-go theorem that prohibits construction of lattice gauge theories for chiral matter with hyperlocal Hamiltonians. \fi 
\iffalse However this is in line with the understanding that lattice gauge theories for chiral fermions can be constructed by relaxing ultralocality \cite{bietenholz1997progress,draper2006locality}. \fi In our model, the sign of effective gap  $W'$ does not change unless $\kappa$ changes sign. However, $\text{sign}(\xi)$ is controlled by the strength of the Zeeman field \cite{supp.}. In the AFM case, $\xi$ changes sign at $h_{c1}.$ Surprisingly, the sign change of the CS level (chirality) comes not from band inversion but  reversal of a velocity component \cite{nasu2018successive}.
 
 In the remaining part of the paper, we study the nature of the topological phases in different field regimes. 
For this purpose we go over to the continuum limit. The lattice gauge field $A_e$ is a line integral of the vector potential $\int_{e} \vec{A}.d\vec{l}.$ The lattice sum passes over to the continuum limit as $\sum _{\text{links}}=\frac{1}{V} \int d^2 x,$
where $V=3\sqrt{3}a^2/2$ is the unit cell volume. The continuum limit CS theory has two abelian $Z_2$ gauge fields and effective (field-dependent) level matrix,
\begin{align}
 K & = \begin{bmatrix}
 \text{sign}(\xi)\,\text{sign}(W') & 1 + \text{sign}(\xi)\,\text{sign}(W') \\ 
 1 + \text{sign}(\xi)\,\text{sign}(W') & 0 \\ 
 \end{bmatrix}.
 \label{eq:Keff}
\end{align}
At low fields ($0 \leq h < h_{c1}$), $\text{sign}(\xi)\,\text{sign}(W')=1,$ so $K_{\text{low}} = 
\begin{bmatrix}
 1 & 2 \\ 
 2 & 0 \\ 
\end{bmatrix},$ which is different from a $Z_2$ spin liquid for which $K_{Z_2} = \begin{bmatrix}
 0 & 2 \\ 
 2 & 0 \\ 
\end{bmatrix}.$ In the AFM case for $h_{c1} \leq h < h_{c2},$ we have $\text{sign}(\xi)\,\text{sign}(W')=-1$ and $K_{\text{inter}} = 
\begin{bmatrix}
 -1 & 0 \\ 
 0 & 0 \\ 
 \end{bmatrix}$. At even higher fields $h>h_{c2}$ corresponding to the partially polarized phase, $\text{sign}(\xi)\,\text{sign}(W')=0$ and we revert to the bare trivial $K$-matrix. For the FM case, single phase transition occurs  at some $h_c,$ from $K_{\text{low}}$ to the trivial one. Having obtained the level matrices in the different field regimes, we now describe the nature of these phases.

Consider first $Z_2$ topological order with level $K_{Z_2}.$
The (abelian) quasiparticles ($e,m,\epsilon$) have the general form $X = e^{i\int_C l_{X}^{\alpha}A_{\alpha \mu}dx^{\mu}}$ where $l^{\alpha}_{X} \in {Z},$ $A_{1\mu}=A_{\mu}$ and $A_{2\mu}=a^{*}_{\mu}.$  
\iffalse
Two different strings anticommute when they cross odd number times as a result of the commutation relations $[A_{\mu}(x),a^{*}_{\nu}(x')]=2\pi i \epsilon_{\mu\nu} K^{-1}_{Z_2}\delta^2(x-x')=\pi i \epsilon_{\mu\nu} \delta^2(x-x').$  
\fi
The quasiparticle charges are $l_e = (1\,\,0)^{T},$ $l_m = (0,\,\,1)^{T},$ and $l_{\epsilon} = (1,\,\,1)^{T}.$
The  fusion rules and braiding statistics follow from the canonical commutation relations $[A_{\mu}(x),a^{*}_{\nu}(x')]=2\pi i \epsilon_{\mu\nu} K^{-1}_{Z_2}\delta^2(x-x')=\pi i \epsilon_{\mu\nu} \delta^2(x-x').$ The Wilson loop $e^{i\oint_{\mathcal{C}} \vec{A}\cdot d\vec{l}}$ gives $+1$ when it encloses  the trivial quasiparticle $\mathbf{1}$ or the $e$ quasiparticle, and $-1$ otherwise. Similarly the Wilson loop $e^{i\oint_{\mathcal{C}} \vec{a^{*}}\cdot d\vec{l}}$ is $+1$ when it encloses  $\mathbf{1}$ or the $m$ quasiparticle, and $-1$ otherwise. The $\epsilon$ particle corresponds to both the eigenvalues being $-1.$ 
Any abelian phase and its topological properties can be completely described by the abelian $K$ matrix. But that is not true in general for any non-abelian phase like Ising topological order (ITO) which we will describe now. For $K=K_{\text{low}},$  $Z_2$ order is clearly lost since the above fusion rules are no longer satisfied.

 For an anyon model which is symmetric under some permutations of their topological charges (e.g. $e\leftrightarrow m$ for $Z_2$), one can describe a non-abelian phase by introducing twist defects \cite{bombin2010topological,teo2015theory,you2013synthetic,bombin2008family} such as dislocations. 
 This is related to a local symmetry operation $G=\begin{bmatrix}
 0 & 1 \\ 
 1 & 0 \\ 
 \end{bmatrix}$ on $K_{Z_2}$ that amounts to interchanging the two species of gauge fields locally. For $Z_2$ topological order, it is known that such twist defects ($\sigma$) are nonabelian ITO anyons \cite{bombin2010topological,zheng2015demonstrating}.
 
\begin{figure}
\centering
\includegraphics[width=\columnwidth]{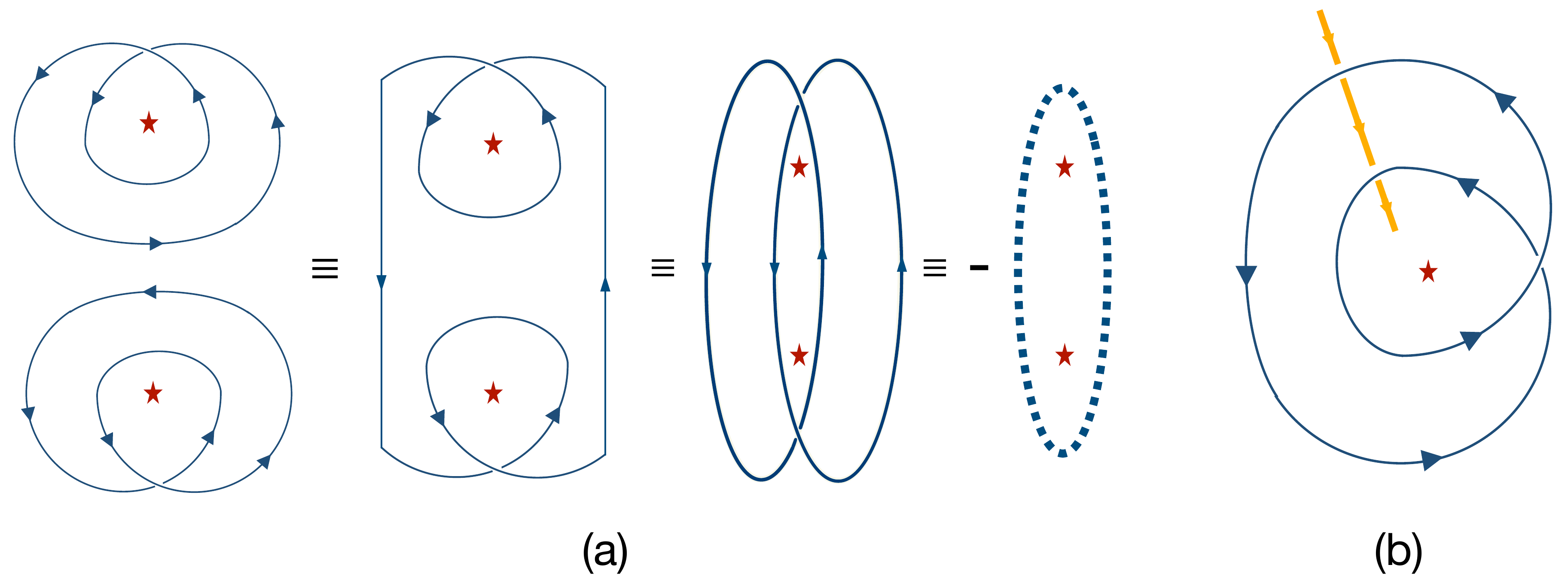}
\caption{\label{fig:fusion_3} Illustration of fusion rules for twist defects. The blue and yellow line represent the string $ e^{i\int\vec{a^{*}}.d\vec{l}}$ and $ e^{i2\int\vec{a^{*}}.d\vec{l}}$ respectively. In (a) fusion of two closed strings that wind twice are shown. The dotted line measures the fermion number parity within. The fusion between the quasiparticles  $\psi$ and $\sigma$ is shown in (b).}
\end{figure}

 Now we show that the structure of $K_{\text{low}}$ naturally admits a twist defect that satisfies the ITO fusion rules -- specifically, the diagonal entry in $K_{\text{low}}$ introduces a single twist that breaks $e$ and $m$ exchange symmetry. We show that the Wilson loop $\Gamma=e^{i\oint_{c_2}\vec{a^{*}}\cdot d\vec{l}},$ where the closed loop $c_2$ winds twice (say anticlockwise) with only one crossing plays the role of detector of a twist defect $\sigma.$ This is analogous to the double winding string detectors of twist defects $\sigma_{\pm}$ in $Z_2$ spin liquids \cite{bombin2010topological} - the main difference in the $Z_2$ case is the time reversal symmetry allows  two distinct twist defects of opposite chirality, and respective eigenvalues $+ i$ or $-i$ for the corresponding double winding detector. In our case, only one of these eigenvalues is realized because all defects have the same chirality. Mathematically, the imaginary eigenvalue for the double winding loop comes from the non-trivial commutation relation $[a^{*}_{\mu}(x),a^{*}_{\nu}(x')]=2\pi i \epsilon_{\mu\nu}K^{-1}_{\text{low}}\delta^2(x-x')=-\frac{\pi}{2}i \epsilon_{\mu\nu}\delta^2(x-x')$  in our model. On the other hand, because of the presence of the diagonal term, $e$ and $m$ charges cannot be defined consistently. We now obtain the fusion rules of these twist defects. In the absence of other excitations, for two closed strings that wind twice, say $\Gamma_1$ and $\Gamma_2,$ we have  $\Gamma_1\Gamma_2=-1$ in Fig. \ref{fig:fusion_3} (a), where the minus sign comes from the presence of two nontrivial crossings, and the remaining dotted curve counts the fermion number parity within. The $\psi$  fermionic quasiparticle can  be realised with the string, $ e^{i 2\int \vec{a^{*}}\cdot d\vec{l}},$ and introducing a $\psi$ into any double winding loop of the type $\Gamma$ simply changes the fermion number parity. The fermion number parity can be $\pm1 $ i.e., modulo $2,$ the charge within is $1$ or $\psi.$  In Fig. \ref{fig:fusion_3} (b), introduction of a $\psi$ (yellow line) into $\Gamma$ clearly does not change the sign, consistent with the fusion rule $\sigma \times \psi =\sigma.$ 
 
 Now we define the Majorana fermion by an open string $ C_j = e^{i\int \vec{a^{*}}\cdot d\vec{l}}$ that loops around a twist defect at $j$ and all Majorana strings thus defined are to have the same end points. Two such strings (say, $C_j$ and $C_{j+1}$) with same end points, when braided, results in the change in the orientation of their intersection, yielding the braiding relations $C_{j}\rightarrow C_{j+1},\,C_{j+1}\rightarrow -C_{j},$ consistent with Refs. \cite{ivanov2001non,bombin2010topological}. The total charge upon fusion is $-iC_j C_{j+1}$, taking values either $+1$ for \textbf{1} or $-1$ for $\psi$ respectively.
  The anyon sector $\{\textbf{1},\psi,\sigma\}$ gives ITO.

 In anyon-CFT correspondence \cite{read1999beyond,read1992fractional,moore1991nonabelions,chen2017orbifolding}, it is well known that these   twist fields in $1+1$ dimensions are either holomorphic or antiholomorphic, which is the case for our model. From the edge theory one can also calculate the conformal weight of these fields and the braiding statistics of ITO. The GSD on the torus for this ITO phase is three due to three superselection sectors discussed above.  The topological entanglement entropy ($\gamma$) can be expressed as $\gamma = \ln D$ where $D$ is the total quantum dimension.
In case of the ITO, $D^2=\sum_{\alpha} d^2_{\alpha}$ with $d=1 $ for vacuum and $\psi$ and $d=\sqrt{2}$ for quasi-particle $\sigma,$ hence $\gamma=\ln 2$ in ITO as expected. On the other hand, the field regime $h_{c1} \leq h < h_{c2}$ corresponds to a chiral abelian phase with trivial topological order and fermionic bulk excitations, so $D=1,\,\gamma=0.$ Note that in the ITO phase, strings composed of the gauge field $A_{\mu}$ do not have any nontrivial commutation relation with the $\sigma$ or $\psi$ quasiparticles of ITO. On the other hand, in the intermediate phase of AFM Kitaev model, they are the sole gauge fields, and give the parity and time reversal symmetry breaking CS term in the theory. The gauge sector gives nontrivial contribution to the central charge in this regime, and is associated with the nonzero vison Chern number. It is understood \cite{moore1991nonabelions,ivanov2001non} that  the  $p+ip$ superconductor supports intrinsic non-abelian Ising anyons upon gauging  the fermion number parity. In our theory, the fermion number parity is associated with the $a^{*}$ dynamical gauge field and $A$ is associated with the vortex number parity. The intermediate field phase is deconfined through $A$ dynamical gauge field which is associated with vortex number parity. 

\textit{Chiral central charge}: From the bulk-edge correspondence \cite{wen1995topological,witten1989quantum,francesco2012conformal},  the effective edge theories of $K_{\text{low}}$ imply the following Kac-Moody algebra of chiral bosons ($\phi$), $ [\phi_{I}(x),\partial_{y}\phi_{J}(y)]=\pm 2\pi i {K^{-1}_{\text{low}}}_{I,J}\delta(x-y).$ 
The fact that number of positive and negative eigenvalues are equal, results in  same number of left-movers and right-movers on the edge, corresponds to zero net chirality ($c=0$). Thus the central charge  in the ITO phase, coming solely from bare chiral Majoranas is $\frac{1}{2}$ \cite{francesco2012conformal}. On the other hand, the intermediate phase of AFM Kitaev model, has an effective  abelian CS theory of level $k=-1$ and leads to  $c=-1$ from the edge theory \cite{francesco2012conformal}, while for the bare chiral Majoranas, the Chern number (-1) is flipped w.r.t. the low field phase \cite{nasu2018successive}. The  resultant chiral central charge  in this phase taking into account both the Majorana and gauge sectors is $c = c_{\text{Majorana}} - c_{\text{gauge}} = \frac{1}{2}.$ 
\iffalse
Remarkably, this topologically trivial phase has the same $c=\frac{1}{2}$ as the ITO phase at lower fields, implying that the half-quantized thermal Hall effect persists with no change in sign, which contradicts the expectation of a change of sign of the thermal Hall effect \cite{nasu2018successive} purely from the parton sector.  
\fi

\textit{Discussion}: In summary, we presented a microscopic derivation of abelian lattice Chern-Simons gauge theories for the FM and AFM Kitaev models subjected to time reversal symmetry breaking Zeeman (001) and scalar spin chirality perturbations. We obtained a comprehensive understanding of the topological phases and emergent excitations in different field regimes. In the low field phase, we proved the long proposed ITO by constructing the intrinsic nonabelian $\sigma$ anyons as twist defects with a specific chirality. In the AFM case, the intermediate field phase associated with trivial topological order was found to be chiral  with $c=1/2,$ (same as the ITO phase, with no sign change) and the same half-quantized thermal Hall response as the ITO phase (see SM \cite{supp.} and Ref. \cite{kapustin2020thermal} for details). Experimentally, the transition to the intermediate phase could be detected as a sharp drop in the plaquette flux value from unity \cite{supp.} but lacking the 1D excitations of the polarized high field phase. Our technique provides a way for understanding other open questions, e.g. the possibility of field-revived ITO in the simultaneous presence of competing ITO degrading perturbations such as Zeeman and exchange interactions.    

Our CS fermionization approach is generally applicable to 2D spin-$1/2$ systems with fermion number parity symmetry, and could be useful in understanding the effects of perturbations on the stability and phase transitions in different spin liquid systems. From a practical perspective, our approach could be useful for realizing fault tolerant nonabelian qubits in an abelian gauge theory setting.

\begin{acknowledgments}
JD and VT acknowledge support of the Department of Atomic Energy, Government of India, under Project Identification No. RTI 4002, and Department of Theoretical Physics, TIFR, for computational resources. AK thanks the National High Magnetic Field Laboratory, which is supported by National Science Foundation Cooperative Agreement No. DMR-2128556* and the State of Florida. AK was supported through a Dirac postdoctoral fellowship at NHMFL.
JD and VT thank specially Subir Sachdev for useful comments on an earlier version of this work, and  Shiraz Minwalla, Ashvin Vishwanath, Yuan-Ming Lu, and Nandini Trivedi for fruitful  discussions.
\newline \newline
\textit{Statement of author contributions}:  JD and VT conceived the problem and performed the theoretical analysis. AK performed the numerical calculations. JD and VT wrote the paper with inputs from AK.
\end{acknowledgments}

%\end{acknowledgments}
%\bibliographystyle{apsrev4-2}
%\bibliography{references.bib}
\bibliographystyle{apsrev4-2}
\bibliography{references}
\end{document}